\documentclass[12pt]{article}
\begin{document}
\begin{center}
 ~~~~~~~~~~~~~~~~~~~~~~~~~~~~~~~~~~~~~~~~~~~~~~~~~~~Preprint INR-TH-2023-016 \\[15mm]

{\huge \bf A New Solution to the Strong CP Problem} \\[10mm]
 S.A. Larin \\ [3mm]
 Institute for Nuclear Research of the
 Russian Academy of Sciences,   \\
 60-th October Anniversary Prospect 7a,
 Moscow 117312, Russia
\end{center}

\vspace{30mm}
\begin{abstract}
We suggest a new solution to the strong CP problem.
The solution is based on the proper use of the boundary
conditions for the QCD generating functional integral.
We expand the perturbative  boundary conditions
to both perturbative and nonperturbative  fields
integrated in the QCD generating functional integral.
It allows to nullify the CP odd term in the QCD Lagrangian
and thus to solve the strong CP problem.
The presently popular solution to the strong CP problem
with introducing axions violates the principle of renormalizability of Quantum
Field Theory, which is very successful phenomenologically.
Our solution obeys the principle of renormalizability of Quantum
Field Theory and
does not involve new exotic  particles like axions.
\end{abstract}

\newpage
\section{Introduction}
The strong CP problem for a long time is considered as in fact an unsolved,
or at least not completely satisfactorily solved,
 outstanding problem
of Quantum Field Theory and Elementary Particle Physics. For an excellent
review of the subject see \cite{gs}, where one can find various aspects of this problem.
The most popular presently solution \cite{pq}, \cite{pq1} to the strong CP problem introduces
new  particles - axions \cite{wa},\cite{wi}. Axions became so popular that
they   are even  considered
as real candidates for dark matter of the Universe.
But presently  only restrictions on their possible properties are established
in spite of the numerous experimental efforts to discover such exotic
particles, see e.g. \cite{expe},\cite{exp1},\cite{exp2}.
Besides, the axion solution of the strong CP problem violates the fundamental principle
of renormalizability of Quantum Field Theory. This basic principle
presently is one of the most phenomenologically successful principles
of Elementary Particles Theory. For example, it ensured 
in Quantum Electrodynamics the agreement between the theory
and the experiment for anomalous magnetic moment of the electron
within  ten decimal points. This impressive agreement
convinces us that renormalizable Quantum Field Theory is a correct 
physical theory.
Therefore, in our opinion,  it seems to be interesting to find
a solution to the strong CP problem which also obeys the principle
of renormalizability of Quantum Field Theory. Besides it is desirable
to find a solution which does not introduce new exotic particles
like axions.
This is the goal of the present paper. 

To find such a solution we will
use in a proper way (a proper way in our opinion) the boundary conditions
in the generating functional of Green functions of Quantum Chromodynamics.
It is well established what kind of
boundary conditions are imposed on the fields
of the theory in the functional integral within perturbative  approach.
These are known boundary conditions which produce the correct form
of the perturbative propagators of the fields of the Lagrangian  in the considered  theory.
The derivation of the perturbative propagators using the boundary conditions
of the generating functional of Green functions of Quantum Chromodynamics
can be found in \cite{fs}.
We will assume that the same boundary conditions are valid for all
fields of the theory which are integrated in the functional integral.
That is we will suppose that perturbation theory calibrates the whole
nonperturbative functional integral.

Our solution  will obey the principle of renormalizability of Quantum
Field Theory and will
not involve new exotic  particles like axions.

The CP problem is the question of why the strong interaction does not violate the
charge-parity (CP) symmetry, which is the combination of charge conjugation (C) and
parity (P) symmetries. The CP symmetry states that the laws of physics should be the
same if a particle is replaced by its antiparticle and its spatial coordinates are inverted
(P). However, the weak interaction is known to violate the CP symmetry, and there is
no fundamental reason why the strong interaction should not do the same.

The CP problem is also related to the origin of the matter-antimatter asymmetry
in the universe, which is another unsolved mystery in physics. The CP violation
involves scalar fields that couple to the quarks and induce a complex phase in the quark
mass matrix. This phase could affect the properties of the neutron stars and black holes
in X-ray binaries, such as their mass, radius, magnetic field, and spin, see e.g.  \cite{evol}. Another possible
connection is that some models of CP violation involve new particles that have spin1/2 and 
interact with the standard model particles via a new force \cite{fuetal}.
These particles could affect the X-ray spectrum or the gravitational
waves emitted by the system.

The CP violation in the early universe could have generated primordial magnetic
fields that were amplified by the collapse of stars into neutron stars. These fields could
then explain the existence of magnetars powered by extremely strong magnetic fields, see
e.g. \cite{chi}. However, this scenario is highly
speculative and requires more theoretical and observational support.

\section{Materials and Methods}
In the present work we will deal with the Quantum Chromodynamics (QCD) generating functional
of Green functions, which will be the basic object of our considerations

\begin{equation}
\label{gf}
Z(J)=\int d\Phi~~ exp\left(i\int d^4x\left( L_{QCD} +J_k\cdot \Phi_k 
\right)\right), 
\end{equation}
where $d\Phi$ denotes the integration measure of the functional integral
$Z(J)$ over all fields $\Phi_k$ of the theory, gluons and quarks.
$J_k$ are sources of the fields. The symbol $J$ in $Z(J)$ denotes
the full set of sources $J_k$ of the fields.

Within  perturbation theory the QCD Lagrangian  $L_{QCD}$ is invariant,
in particular, under the combined symmetry transformations CP, where C is the
charge conjugation operator and P is the space reflection. More precisely
the QCD Lagrangian within perturbation theory is invariant under both the charge conjugation C and
the space reflection C.
 
The essence of the CP problem is that in full nonperturbative QCD
 one can add to the QCD Lagrangian
the CP odd gauge invariant term which seems to be  not forbidden from the first
principles

\begin{equation}
\label{delta}
\Delta L_{\theta}=\frac{\theta}{32\pi^2}G_{\mu\nu}^a \tilde{G}_{\mu\nu}^a.
\end{equation}
It is invariant under the charge conjugation C  and
is not invariant under the space reflection P, hence it is also
non invariant under the combined  CP transformation.
But this term is forbidden by experiments
with the rather high precision as we will see it below.

The dual field strength tensor $\tilde{G}_{\mu\nu}^a$ in (\ref{delta})
is defined in the standard way

\begin{equation}
\label{dual}
\tilde{G}_{\mu\nu}^a=\frac{1}{2}\epsilon_{\mu\nu\rho\sigma}G_{\rho\sigma}^a.
\end{equation}
The $\theta$-term in (\ref{delta}) is purely nonperturbative 
since it is invisible in perturbation theory because
it can be rewritten as a total derivative

\begin{equation}
\label{total}
\Delta L_{\theta}=\theta\partial_{\mu}K_{\mu}.
\end{equation}
Here $K_{\mu}$ is the known Chern-Simons current

\begin{equation}
\label{dual1}
K_{\mu}=\frac{1}{16\pi^2}\epsilon_{\mu\nu\rho\sigma}\left(
A_{\nu}^a\partial_{\rho}A_{\sigma}^a+\frac{1}{3}f^{abc}A_{\nu}^aA_{\rho}^b A_{\sigma}^c\right).
\end{equation}
The $\theta$-term can be discarded within perturbation theory.
It can be easily seen in the Euclidean space
since the fields of the theory decrease in the Euclidean space at the time infinities and
the total derivative (\ref{total}) does not contribute to the QCD action.
But with the discovery of instantons \cite{inst} it was realized
that the field configurations with the instanton boundary conditions
give nonzero nonperturbative contributions to the action.
In particular, the one instanton contribution looks like

\begin{equation}
\label{action}
\Delta S_{\theta}=\int d^4x \Delta L_{\theta}=\theta.
\end{equation}
The key notion here is the famous topological charge which
has the following form

\begin{equation}
\label{top}
\mathcal{V}=\int d^4x \partial_{\mu}K_{\mu}=\int d^3x K_0(\vec{x},t)|^{t=+\infty}_{t=-\infty}
=\mathcal{K}(t\rightarrow+\infty)-\mathcal{K}(t\rightarrow -\infty),
\end{equation}
where $\mathcal{K}$ is the Pontryagin number.
The topological charge is zero for perturbative fields, i.e.
in perturbation theory. But instanton fields, for example in the $A_0=0$
gauge, interpolate between the zero gluon fields $A_i(\vec{x},t=-\infty)=0,~~i=1,2,3$
and the nonzero  gluon fields $A_i(\vec{x},t=+\infty)=U^+\partial_iU,~~ i=1,2,3$.
Here the matrix $U$ is the Polyakov hedgehog

\begin{equation}
\label{hedg}
U(\vec{x})=exp\left(-\frac{i\pi\vec{x}\cdot\vec{\sigma}}
{\sqrt{\vec{x}^2+\rho^2}}\right).
\end{equation}
For this instanton configuration one has that the Pontryagin number
and correspondingly the topological charge are equal to unity
 
\begin{equation}
\mathcal{V}=\mathcal{K}(t=+\infty)=1, 
\end{equation}
Thus the $\theta$-term gives the nonzero nonperturbative contribution to the QCD action.

In the full QCD, with quarks, there is also  contributions to the
CP odd part of the QCD Lagrangian from the imaginary phases of the quark
mass matrix. The phases can be rotated away by the chiral transformations
of quark fields. But there is the famous axial anomaly \cite{ax},\cite{ax1}. 
It generates
non invariance of the measure of the Feynman functional integral under
chiral transformations \cite{f}. Therefore the phases of the quark
mass matrix arise before the $G\tilde{G}$ term in the Lagrangian.
Hence the parameter which determines the value of the CP violation is in fact

\begin{equation}
\theta + arg (det \mathcal{M}),
\end{equation}
where $ \mathcal{M}$ is the quark mass matrix.

Below we shall use the same symbol $\theta$ for this parameter
to simplify notations assuming that it already includes the effects
of the quark mass matrix. 

Probably the most essential phenomenological effect of the $\theta$-term is a nonzero
electric dipole moment of the neutron $d_n$. The electric dipole moment is given by the 
effective interaction Lagrangian

\begin{equation}
L_{nEDM}=\frac{d_n}{2}\bar{n}i\gamma_5\sigma_{\mu\nu} n F^{\mu\nu},
\end{equation}
where $F^{\mu\nu}=\partial^{\mu}A^{\nu}-\partial^{\nu}A^{\mu}$ is the 
photon field strength tensor, $n$ stands for the neutron field ,
$\sigma_{\mu\nu}=\frac{1}{21}[\gamma_{mu}\gamma_{nu}]$
is the standard antisymmetric product of Dirac gamma matrices.

The $\theta$-term generates the following electric dipole moment of the neutron $d_n$

\begin{equation}
\label{dn}
< n(p_f)\gamma(k)|e J_{\mu}^{em}A^{\mu}i\int d^4x\Delta L_{\theta}|n(p_i)>=
d_n \bar{n}(p_f)\gamma_5\sigma_{\mu\nu}n(p_i)k^{\mu}\epsilon^{\mu}(k).
\end{equation}
Here $J_{\mu}$ is the quark electromagnetic current. 
$k^{\mu}=p_f^{\mu}-p_i^{\mu}$, 
where $p_i^{\mu}$ is the incoming momentum of the neutron
and $p_f^{\mu}$ is the outgoing momentum of the neutron.
$\epsilon^{\mu}(k)$ is the photon polarization.

The matrix element in the left hand side of eq.(\ref{dn}) is zero in perturbative
theory and is calculated
within purely nonperturbative QCD. There are several nonperturbative  methods  by which
the electric dipole moment of the neutron $d_n$ was
estimated, for the detailed overview see  \cite{gs} and references therein. 
Here we give the short summary of the results of the corresponding
nonperturbative approaches.
The performed bag model calculations produced the following result 
$d_n \approx \theta 2.7\cdot 10^{-16}$ e$\cdot$cm. 
Shortly after this result the chiral logarithms approach was used to obtain
the following estimate 
$d_n \approx \theta 5.2\cdot 10^{-16}$e$\cdot$cm.
The approach of chiral perturbation theory was further developed to produce
the slightly less result 
$d_n \approx \theta 3.3 \cdot 10^{-16}$e$\cdot$cm.
At last but not at least the calculations based on the QCD sum rules method
gave the following again slightly less estimate 
$d_n \approx \theta 1.2\cdot 10^{-16}$e$\cdot$cm.
All these results  have considerable uncertainties
of the order of $50$ per cent because of essential difficulties of nonperturbative
QCD calculations. 

But anyway the average theoretical value for $d_n$
can be confidently estimated within the same of the order of $50$ per cent uncertainty as

$d_{n,theor}\approx \theta\cdot 10^{-16}$e$\cdot$cm.

This should be compared with the most recent experimental value \cite{ed} for 
the electric dipole moment of the neutron $d_n$
which is 

$d_n=(0.0\pm 1.1)\times10^{-26}$e$\cdot$cm.

Thus  one gets an extremely strong restriction on the value of the $\theta$ coupling

\begin{equation}
|\theta| \le 10^{-10}.
\end{equation}
The explanation of this  practically zero value
 of the coupling $\theta$
is the essence of a solution to the strong CP problem.

The presently popular solution to the problem is the famous axion solution.
It assumes the addition to the QCD Lagrangian the term with the new
axion field $a(x)$, which in fact reduces to the shift of the coupling $\theta$
in the QCD Lagrangian 
$\theta\rightarrow a(x)/f_a+\theta$. So the corresponding term
$\Delta L_{\theta}$ of eq.(\ref{delta}) in the
QCD Lagrangian becomes as follows

\begin{equation}
\label{ax}
\Delta L_{\theta}\rightarrow \left(\frac{a(x)}{f_a}+\theta \right)
\frac{1}{32\pi^2}G_{\mu\nu}^b \tilde{G}_{\mu\nu}^b.
\end{equation}
After the spontaneous symmetry breaking of the global Peccei Quinn symmetry \cite{pq},\cite{pq1} 
one calculates the effective potential for the axion field $a(x)$. Then
one finds that when the axion rests at the minimum of this potential,
the CP violating term (\ref{ax}) nullifies. This is the known axion solution to 
the strong CP problem.

There are different types of axions suggested in the literature. 
Let us consider some of them.
As the scalar Higgs fields produce vacuum expectation values the electroweak
local symmetry group is broken spontaneously. This develops masses of the 
gauge W and Z intermediate vector bosons. At the same time the global 
Peccei Quinn U(1) group is also spontaneously broken.This spontaneous breaking
of the U(1) global symmetry leads to the appearance of the massless Goldstone
boson, which is called the Weinberg-Wilczek (WW) axion in this case \cite{wa},\cite{wi}.
In the Standard Model including two Higgs doublets this axion is presented
as the following superposition

$a=1/v(v_{\phi}Im\phi_0-v_{\chi}Im\chi_0)$,

here $\phi_0$ and $\chi_0$ are neutral components of the two Higgs doublets.
Besides, $v=\sqrt{v_{\phi}^2+v_{\chi}^2}\approx$ 250 GeV, where $v_{\phi}$ and
$v_{\chi}$ denote  vacuum expectation values of the fields $\phi$ and $\chi$
correspondingly. Within the considered approximation the WW axion is massless.
But as it was mentioned above the nonperturbative effects of
Quantum Chromodynamics (for example instantons) can generate the potential
for the WW axion. In this way the WW axion obtains the nonzero mass value
which is estimated according to \cite{wa},\cite{wi} as following

$m_a\approx f_{\pi}m_{\pi}/v \approx$ 100 KeV.

Besides, the decay constant of the WW axion is $1/v$. Hence it is clear 
that the mass and the decay constant of the WW axion is connected to the
breaking scale $v$ of electroweak symmetry. This constraint turns out to be
too strong and correspondingly the WW axion turns out to be excluded by the 
existing experimental data.

If the scale of the breaking of the Peccei Quinn symmetry is much more than
the electroweak scale $v$, then according to the above formula the axion is 
essentially lighter and the decay constant of the axion is much smaller.
This type of the light axion could be in agreement with the existing 
experimental data.

A solution with the light 'invisible' axion was first suggested in \cite{ku},
\cite{sv} (the so called KSVZ axion). In the Ref. \cite{sv} this type of 
the axion is named the phantom axion. It should be underlined that in order to
uncouple the 'phantom' from the electroweak scale it is necessary to decouple
the proper scalar fields from the standard quarks and couple these fields to 
very heavy hypothetical fermions carrying color.

To be more precise one should introduce  a complex scalar $\Phi$ which 
is coupled to the hypothetical quark field $Q$, the electroweak singlet in
the fundamental representation of the SU(3) color group.

Then the modulus of the scalar $\Phi$ is supposed to produce the large
vacuum expectation value $f/\sqrt{2}$ and the argument of the field $\Phi$
is just the axion field $a$ up to normalization:
\[ a(x)=f \alpha(x),~~~\alpha(x)= Arg\Phi(x),~~~~ f>>\Lambda.\]

Further the low energy coupling of this axion to the gluons is as follows
\[ \Delta L=\frac{1}{f}a\frac{1}{32\pi^2} G_{\mu\nu}^a \tilde{G}_{\mu\nu}^a.\]
In this way the Lagrangian of QCD depends on the expression $\theta+\alpha(x)$.

More generally, it is possible to introduce more than one quark field $Q$
or to introduce these fields in a higher representation of the SU(3)
color group. In this case the coupling of the axion to gluons obtains
an integer number $N$:
\[\Delta L=\frac{1}{f}aN\frac{1}{32\pi^2} G_{\mu\nu}^a \tilde{G}_{\mu\nu}^a.\]
This multiplier $N$ (should not be confused with the color number $N_c$) is 
usually called as the axion index. Hence, in general, the Lagrangian of 
Quantum Chromodynamics depends on the sum $\theta +N \alpha(x)$. It can be 
assumed that
the nonperturbative QCD effects produce the potential for 
$\theta +N \alpha(x)$. The latter sum is minimized at the point
$\theta +N \alpha_{vac}=0$. Hence the strong CP problem is solved in this way.

One more way to produce an 'invisible' axion is suggested 
in \cite{zh},\cite{df} (the so called ZDFS axion). In this case one keeps
the Peccei Quinn symmetry of the two doublet Standard Model but splits 
the scales of the Peccei Quinn and electroweak breaking.
For this purpose the Lagrangian of the Standard Model  is extended, one adds
the scalar Standard Model singlet field $\Sigma$

Then one note that this Lagrangian is in variant under the axial 
transformations
\[q_L\rightarrow e^{i\alpha}q_L,~~q_R\rightarrow e^{i\alpha}q_R, ~~
\phi\rightarrow e^{2i\alpha}\phi,~~ \chi \rightarrow e^{-2i\alpha}\chi,~~
\Sigma \rightarrow e^{2i\alpha}\Sigma.\]
After spontaneous breaking of this axial symmetry the Goldstone particle, 
the axion, appears as the superposition

\[a=\frac{1}{V}(v_{\phi}Im\phi_0-v_{\chi}Im\chi_0+v_{\Sigma}Im\Sigma.\]
Here $V=\sqrt{v_{\phi}^2+v_{\chi}^2+v_{\Sigma}^2}$, $v_{\phi},~v_{chi}$ and
$ v_{\Sigma}$ represent vacuum expectation values of the fields $\phi,~\chi$
and $\Sigma$ correspondingly.
The vacuum expectation value of the field $\Sigma$ is not necessarily connected
to the scale of the electroweak symmetry breaking.
Actually it can be chosen as large as the Grand Unification scale.
In this case the considered axion is very light and its decay constant 
is small.

But first of all, the term with the axion field $a(x)$ in (\ref{ax})
has the dynamical dimension (that is the dimensions of the fields plus
the dimensions of the derivatives of the fields) five instead of four necessary for 
renormalizability. Hence the term with the axion field $a(x)$ violates renormalizability
of the Lagrangian. As we already have mentioned in the introduction,
renormalizability of the Lagrangian is a rather important principle of Quantum Field Theory.
This principle turned out to be  very successful phenomenologically as it is demonstrated
for example by the famous case of the anomalous magnetic moment of the electron. Therefore it is quite
important to preserve the principle of renormalizability when solving the strong CP problem.
Secondly, the axion is not found experimentally in spite of
numerous experimental attempts, as we also have underlined in the introduction.

Therefore we find it necessary to suggest a new solution
to the strong CP problem which preserves renormalizability of the theory
and does not involve new exotic particles like axions.

Let us now again consider the QCD generating functional (\ref{gf}).
It is well known that this integral is not defined yet completely if
only the QCD Lagrangian is defined with the corresponding gauge condition.
At least within perturbation theory one should impose on the Lagrangian fields
the proper boundary conditions.

In perturbation theory one has the well known boundary conditions. For example
for the gluon fields one has the following conditions
\begin{equation}
\label{bou}
A_{\mu}^a(\vec{x},t \to - \infty) \to A_{\mu,in}^a(x),
\end{equation}
\[
A_{\mu}^a(\vec{x},t \to + \infty) \to A_{\mu}^{a,out}(x). 
\]
Here the incoming asymptotic gluon  fields $A_{\mu,in}^a(x)$ contain only
the positive frequency part and the  outgoing gluon  fields $A_{\mu,}^{a,out}(x)$
contain only the negative frequency part:

\begin{equation}
\label{boun}
A_{\mu,in}^a(x)=\frac{1}{(2\pi)^{3/2}} \int d^3 k~ e^{i(\vec{k}\vec{x}-\omega t)}
v_{\mu}^i(k)a_i(k)/\sqrt{2w}, \\
\end{equation}
\[
A_{\mu,}^{a,out}(x)=\frac{1}{(2\pi)^{3/2}} \int d^3 k~ 
e^{i(\vec{k}\vec{x}+\omega t)}
v_{\mu}^i(k)a^*_i(k)/\sqrt{2w},
\]
where $\omega=\sqrt{\vec{k}^2}$ and $v_{\mu}^i(k)$  are
polarization vectors of the gluons. Here the sums over
the gluon polarizations $i=1,2$ are assumed.

These are the known Feynman boundary conditions. They
are necessary to obtain the correct form of the perturbative propagators
of the fields
of the type $1/(k^2+i\epsilon$) with the correct plus $i\epsilon$ prescription.
Thus with this boundary conditions the gluon fields (the quark fields also)
 oscillate at the time infinities. After transition to the 
Euclidean space by means of the Wick rotation $t\rightarrow ix_4$ the fields
decrease at the time infinities, so in the Euclidean space it is easy to see that in perturbation theory
total derivatives in the Lagrangian are zero.

Hence one can  write perturbative boundary conditions (\ref{bou}) for all fields
$\Phi_i$ of the  QCD Lagrangian symbolically as follows
\begin{equation}
\label{bou1}
\Phi(t \to \pm\infty)\to\Phi_{in}^{out}(x).
\end{equation}
This helpful notation will be used below to formulate the QCD generating functional
integral as a compact  formula.

\section{Results}
Let us now exactly formulate our solution to the strong CP problem.
As it is was already underlined above the solution is based on the 
proper use of the boundary conditions for all Lagrangian fields
in the QCD generating functional integral.
So let us now again consider the QCD  generating functional integral (\ref{gf}.
  
In our opinion, it is necessary and natural to generalize the  boundary
conditions (\ref{bou1}) for perturbative fields to all
fields $\Phi_i$ of the QCD Lagrangian which are integrated in the generating functional
integral (\ref{gf}). Then all Lagrangian fields will decrease in the
Euclidean space at the time infinities. Hence such a definition of
the boundary conditions  nullifies all
total derivatives in the Lagrangian both for the perturbative
contributions and the nonperturbative contributions. Thus the CP odd term in the QCD Lagrangian 
will be nullified and  it solves the strong CP problem.
Besides this definition allows to formulate exactly the complete (the perturbative
part plus the nonperturbative part) QCD generating functional
integral as one compact mathematical formula:
\begin{equation}
\label{gf1}
Z(J)=\int_
{\Phi(t \to \pm\infty)\to\Phi_{in}^{out}}
 d\Phi~exp\left(i\int d^4x\left( L_{QCD} +J_k\cdot \Phi_k
\right)\right).
\end{equation}

\section{Discussions}
The strong CP problem for a long time is in fact considered as the
still unsolved  (or not completely adequately solved) prominent problem
of Quantum Field Theory and Elementary Particle Physics. For an excellent
review of the CP problem and related topics see the fist reference
of the present paper, where one can discover different aspects of the subject.
The presently  most popular axion solution \cite{pq}, \cite{pq1} to the strong CP problem introduces
new exotic elementary particles - axions. Axions are now so popular that
they   are presently  considered
as the real candidates for  Dark Matter of the Universe. A lot of advanced experiments
are performed to discover some sorts of these axions.
But presently  only some kinds of  restrictions on their possible properties are obtained
in spite of the numerous huge experimental efforts to find such exotic elementary
particles, see for example \cite{expe},\cite{exp1},\cite{exp2}.
Besides, the axion solution of the strong CP problem is in contradiction with 
the fundamental, in our opinion, principle
of renormalizability of Quantum Field Theory. This basic, in our opinion, principle
presently is one of the most experimentally successful principles
of Quantum Field Theory and
Elementary Particles Physics. For example, this principle produced
for the famous anomalous magnetic moment of the electron
within renormalized Quantum Electrodynamics
the outstanding agreement between the theoretical value
and the experimental value 
within  ten decimal points. This prominent agreement between the theory and the experiment
convinces us that renormalizable Quantum Electrodynamics and 
more generally renormalizable Quantum Field Theory are the proper
physical theories. 

Therefore, in our opinion,  it seems to be important to have
a solution to the strong CP problem which is in agreement with the principle
of renormalizability of Quantum Field Theory. Besides we suppose that it is desirable
to find a solution to the strong CP problem which does not involve new exotic elementary particles
like axions or something similar.
Therefor the goal of the present paper is to find the solution which satisfies
these two conditions (renormalizability and the absence of new exotic particles).

To find such a solution we have 
used in this paper in a proper way the boundary conditions for the Lagrangian fields
in the generating functional of the Green functions of Quantum Chromodynamics
(a proper way in our opinion).
It is well understood what kind of
boundary conditions should be imposed on the fields
of the theory in the generating functional integral within perturbation  approach.
These are the known boundary conditions which generate the necessary form
of the perturbative propagators of the fields of the Lagrangian of  Quantum Chromodynamics.
These boundary conditions produce the correct '$+i\epsilon$' prescription
for the perturbative propagators of the fields.
We have suggested that the same boundary conditions should be valid for all
Lagrangian fields   integrated in the 
QCD generating functional integral, i.e. for both perturbative and nonperturbative
contributions.
That is we have supposed that perturbation theory calibrates the whole
nonperturbative functional integral, i.e. it calibrates both the perturbative and
the nonperturbative parts of the generating functional integral.

Our solution satisfies two important criteria. The solution does  obey 
the principle of renormalizability of Quantum
Field Theory and does
not involve new exotic  particles like axions.

Let us make now necessary here remarks concerning the famous  $U(1)$ problem.
The essence of this problem is that the mass of the flavour singlet
pseudo-scalar $\eta'$ meson  $m_{\eta'}\approx 958 MeV$ is 
surprisingly heavier than the masses
of the flavour octet pseudo-scalar mesons.
One can argue that it is not possible to extend the discussed above perturbatve
boundary conditions for the fields, which are well established within
perturbation theory,  to all fields which are integrated in the 
QCD generating functional integral. Such an extension excludes from
the generating functional integral for example the instanton contributions
not obeying the perturbative boundary conditions.
In particular there is the well known statement \cite{h},\cite{h1} that instantons solve
the  $U(1)$ problem. Hence it seems that they should not be excluded
from the theory. But one can note that there is also the well known solution \cite{ks},\cite{ks1}
to the  $U(1)$ problem using the axial anomaly
which was suggested before the discovery of instantons.
Therefore one can argue  that the  $U(1)$ problem can be solved without
involving the instantons. Thus the extension of the perturbative boundary
conditions to all fields of the generating functional integral is well allowed.

\section{Conclusions} We have  suggested a new solution to the  strong CP problem.
To find such a solution we 
use in a new way the boundary conditions
in the Quantum Chromodynamics generating functional of Green functions.
It is well established what kind of
boundary conditions are imposed on the fields
of the QCD Lagrangian in the functional integral within perturbation theory.
We assume that the same boundary conditions are valid for all
Lagrangian fields of the functional integral, i.e.
for both perturbative and nonperturbative fields.
This allows to nullify total derivatives in the QCD action, in particular
the CP odd term which can be presented as the total derivative.
Hence it solves the strong CP problem. 
Thus we suppose that perturbation theory calibrate the complete
nonperturbative functional integral. 

Maybe it is worthwhile also  to mention that our solution does not violate
any symmetries of Quantum Chromodynamics.
We would like to underline once more that our solution to the strong CP problem
obeys the principle of renormalizability of Quantum
Field Theory and
does not involve new exotic  particles like axions.

\section{Acknowledgments}
The author is grateful
to the collaborators of the Theory Division of the Institute 
for Nuclear Research
for valuable discussions.

\end{document}